\documentclass[useAMS,usenatbib,usegraphicx]{mn2e}
\usepackage{psfig}   
\usepackage{graphicx}
\usepackage{subfigure}
\newcommand{\msun}{~M$_\odot$} 

\title[System Mass not Primary Mass]{Binary mass ratios: system mass not primary
mass}

\author[S.~P.~Goodwin]{
  Simon P.~Goodwin\thanks{E-mail: s.goodwin@sheffield.ac.uk}
  \vspace*{0.1cm}\\
   Department of Physics and Astronomy, University of Sheffield, Sheffield, S3 7RH, UK}

\begin{document}

\date{}
                             
\pagerange{\pageref{firstpage}--\pageref{lastpage}} \pubyear{2012}

\maketitle

\label{firstpage}

\begin{abstract}
Binary properties are usually expressed (for good observational
reasons) as a function of primary mass.  It has been found that the
distribution of companion masses -- the mass ratio distribution -- is
different for different primary masses.  We argue that system mass is
the more fundamental physical parameter to use.  We show that 
if system masses are drawn from a log-normal
mass function, then the different observed mass ratio distributions
as a function of primary
mass, from M-dwarfs to A-stars, are all consistent with a universal,
flat, system mass ratio distribution.  We also show that the brown
dwarf mass ratio distribution is not drawn from the same flat
distribution, suggesting that the process which decides upon mass
ratios is very different in brown dwarfs and stars.
\end{abstract}


\section{Introduction}

It has been known for a long time that many stars are physically
associated in binary and multiple systems (Mitchell 1767).  Recent
studies suggest that many, possibly most, stars in the field are in
multiple systems (e.g. Duquennoy \& Mayor 1991; Fischer \& Marcey
1992; Lada 2006; Bergfors et al. 2010; 
Raghavan et al. 2010; Janson et al. 2012; de Rosa et al. 2012).  Numerical experiments have
shown that it is extremely difficult to dynamically produce a binary
in star-forming environments
(Clarke \& Pringle 1991; Kroupa 1995), which suggests that almost all 
binaries form as binaries.  Therefore the properties of 
binaries should contain a significant
amount of information on the star formation process.

Observational binary surveys often take the approach of
selecting a sample of stars of a particular spectral type and
examining them for the existence of companions (e.g. Duquennoy \& Mayor
1991; Fischer \& Marcey 1992; Kouwenhoven et al. 2007; Raghavan et
al. 2010; Bergfors et al. 2010; Janson et al. 2012; de Rosa et
al. 2012).  This is a perfectly sensible observational strategy,
however it means that the results are presented as binary fractions,
separation distributions, mass ratio distributions etc. as {\em 
a function of primary mass}.  In this paper we will show that our
interpretation of binary data can change if we look at distributions
of binary properties by {\em system mass} rather than by primary mass.

Binaries can be described in terms of four basic parameters.  Each
system will have a mass ratio, $q$, between the primary star (mass
$M_{\rm p}$), and secondary (mass $M_{\rm s}$), $q = M_{\rm s}/M_{\rm
  p}$.  The system will also have orbital parameters of semi-major
axis and eccentricity.  Within a population (however that may be
defined) there is also a `binary fraction' that measures what fraction
of that population are binaries (see Reipurth \& Zinnecker 1993).  It
is in the distributions of these properties that we hope to find
information about the star formation process (see e.g. King et
al. 2012a,b).

Most stars spend at most a few Myr in the fairly dense (compared
  to the field) 
star-forming regions in which they are born before dispersing 
into the field (Lada \& Lada 2003).  However, the interpretation 
of binary properties is complicated by the fact that many binaries are
quite easy to dynamically destroy in their relatively dense birth
environments (Heggie 1975; Hills 1975).  In
particular, the binary fractions and semi-major axis distributions of
populations can be significantly altered, and how they are altered
depends strongly on the density of the environment (see Kroupa 1995; King et al. 2012b
and references therein).

Recently Parker \& Goodwin (2012) showed that the mass ratio
distribution of binaries is generally not significantly altered by
dynamics.  This suggests that examining mass ratio distributions is a
way of probing the outcome of star formation without having to account
for the many potential problems of dynamical processing.  Therefore in
this paper we will examine the mass ratio distributions of binary
systems which are often modelled as having the form $f(q) \propto
q^{\alpha}$ where $\alpha >0$ favours more equal-mass systems, and
$\alpha <0$ favours more unequal-mass systems.  

It is known that the binary mass ratio distributions are inconsistent
with random pairing from the IMF (see Kouwenhoven et al. 
2009; Reggiani \& Meyer 2011).  Random pairing from the IMF is
probably not what is expected from physical arguments (see Kouwenhoven et al. 
2009).  However, it is unclear what physics of star formation does set
the mass ratio distributions of binaries.  An interesting
observation is that the mass ratio distribution for G-dwarfs is
independent of separation (Metchev \& Hillenbrand 2009), however it is
completely unclear if this observation extends to other primary masses.

We take four ranges of {\em primary} masses for which there is reasonable
observational data on the mass ratio distributions.  

\noindent 1. Brown dwarf primaries with masses $0.04 < M_{\rm
  p}/$\msun$ < 0.08$.  
Brown dwarfs strongly favour equal-mass
companions (Burgasser et al. 2007).  Examination of the Very Low-Mass
Binary (VLMB) Archive\footnote{Data as of August 2012,
  http://vlmbinaries.org/} shows that of 97 VLMBs, only 7 have
$q<0.6$, and 47 have $q>0.9$.  A very rough fit suggests $\alpha \sim
+1.5$.  Brown dwarfs have a low binary fraction of around 
10--20 per cent (Burgasser et al. 2007).

\noindent 2. M-dwarf primaries with masses $0.1 < M_{\rm
  p}/$\msun$ < 0.5$.  
Samples of field M-dwarf 
binaries have been investigated recently 
by Janson et al. (2012; see also Bergfors et al. 2010).  Janson 
et al. (2012) find from a detailed analysis of
their large sample that an underlying uniform mass 
ratio distribution is most likely, ie. $\alpha \sim 0$ (see their
section~7.1).   M-dwarfs have a binary fraction of around 30--40 per cent 
(Fischer \& Marcey 1992; Lada 2006; Bergfors et al. 2010; Janson et
al. 2012).

\noindent 3. G-dwarf primaries with masses $0.9 < M_{\rm
  p}/$\msun$ < 1.1$.  
The mass ratio distribution of G-dwarfs is
uncertain.  Metchev \& Hillenbrand (2009) find a tendency
to lower-mass companions with $\alpha \sim -0.4$, but Raghavan et
al. (2010) find a roughly uniform distribution with $\alpha \sim 0$
(but with a significant $q \sim 1$ peak).   The binary fraction of
G-dwarfs appears to be 50--60 per cent 
(Duquennoy \& Mayor 1991; Metchev \& Hillenbrand 2007; 
Raghavan et al. 2010).

\noindent 4. A-star primaries with masses $2 < M_{\rm
  p}/$\msun$ < 3$.  
We take two A-star surveys, de Rosa et al. (2012) for the
field, and Kouwenhoven et al. (2007) for Sco OB.  Both find a mass
ratio distribution that favours lower-mass companions with a
distribution roughly $f(q) \propto q^{-0.4}$ (Kouwenhoven et al. 
2007).  It should be noted that
both samples are of relatively distant visual 
companions, typically at several hundred AU.  Shatsky \& Tokovinin
(2002) found a similar result for B-stars in Sco OB with $f(q) \propto
q^{-0.5}$. The binary fraction in A-stars is high at around  $\sim 80$ per
cent in A-stars (Kouwenhoven et al. 2007; de Rosa et al. 2012; Peter 
et al. 2012).

In this paper we first argue that system mass is the underlying
physical distribution in which we should be interested.  We then
perform a simple Monte Carlo experiment picking system mass from a
log-normal system mass function and examining the variation 
mass ratio with primary mass and comparing it with the observations.  

\section{Selecting by system mass}

We argue that to compare binaries one must examine the distributions
of binary properties by system mass rather than by primary mass.  If
most binaries are primordial (as we argued above) then some physical
process acts to fragment (or not) a system into two components and distributes
the mass between the two components\footnote{We will ignore if a
  system fragments into more
  than two objects, however we will discuss higher-order multiplicity
  in the conclusions.}.  Therefore, the fundamental mass
is the system mass $M_{\rm sys} = M_{\rm p} + M_{\rm s}$, not just the mass
of the primary star.  

For example, when comparing three systems of component masses 
(1) 0.6\msun + 0.6\msun,
(2) 1\msun + 0.2\msun, and (3) 1\msun + 0.8\msun we
would argue that it is systems (1) and (2) that should be considered
similar (they both have system mass 1.2\msun), not systems (2)
and (3) (which have primary masses of 1\msun, but very different
system masses).

\subsection{A Monte Carlo experiment}

To see what the differences between selecting by system mass rather
than primary mass can make we perform a simple Monte Carlo experiment.

Let us take the log-normal system mass function with $\mu = -0.7$ and
log-dispersion $\sigma = 0.6$ from Chabrier (2003).  We then 
split each system into a binary with a mass ratio
distribution $f(q) \propto q^{\alpha}$ with a minimum allowable value
of $q = 0.1$.  We also limit the lower-mass of a secondary to 
be 0.01\msun which has the effect of raising the minimum-$q$ for very
low-mass systems.

We will then select binaries by the mass of the primary star and
examine the mass ratio distribution as a function of primary mass.  We
take the four primary mass ranges described above: brown dwarfs of
0.04--0.08\msun, M-dwarfs of 0.1--0.5\msun, G-dwarfs of 0.9--1.2\msun,
and A-stars of 2--3\msun.

It turns-out that the most interesting underlying mass ratio
distribution is the simplest -- when it is flat, ie $\alpha =
0$ always.

\begin{figure}
\begin{center}
   \includegraphics[width=6cm,angle=270]{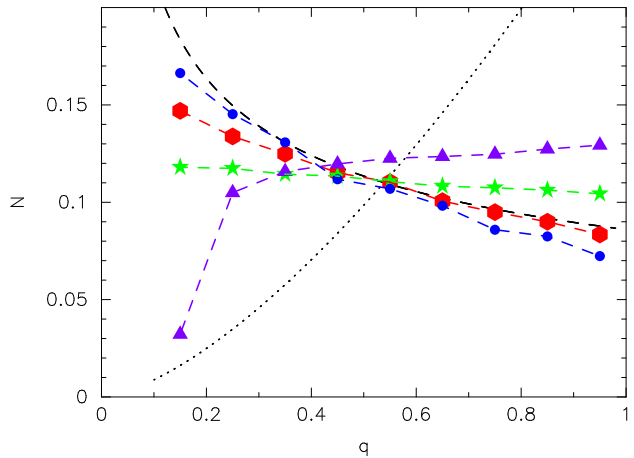}
    \caption{The mass ratio distributions (normalised to unity) of
      systems with brown dwarf (purple triangles), M-dwarf (green
      stars), G-dwarfs (red hexagons), and A-stars (blue circles)
      from an underlying flat system mass ratio distribution.  For
      reference the $f(q) \propto q^{-0.4}$ (black dashed line), and
      $f(q) \propto q^{1.5}$ (black dotted line) distributions are 
      shown.}
    \label{alphazero}
\end{center}
\end{figure}

In fig.~\ref{alphazero} we show
the resulting mass ratio distributions by primary mass.  Brown dwarfs
are shown by purple triangles, M-dwarfs by green stars, G-dwarfs by
red hexagons, and A-stars by blue circles.  For reference
the dashed line shows the distribution for $\alpha = -0.4$ (the observed
value for A-stars), and the dotted line for $\alpha = +1.5$ (a rough 
fit to the brown dwarf observations).

It is clear in fig.~\ref{alphazero} that despite a universal system
mass ratio distribution, the mass ratio distributions by primary mass
are different.  The underlying system mass ratio distribution is flat,
but both A-stars and G-dwarfs show a preference for more unequal-mass
companions, whilst M-dwarfs are flat, and brown dwarfs favour somewhat
more equal-mass companions (note that the low number of brown dwarf
systems in the first bin with $q =0.1$--$0.2$ is mainly due to the 
constraint that $M_{\rm s}$ be greater than 0.01\msun).

The reason for this is that for a binary to have a primary 
mass of $M_{\rm p}$ then the system
must have a mass of $\sim M_{\rm p} < M_{\rm sys} \leq 2 M_{\rm p}$
(formally, if $M_{\rm sys} = M_{\rm p}$ then the system is a single
star).  If $M_{\rm sys} = 2 M_{\rm p}$ then $q = 1$, if $M_{\rm sys}$
is only a little greater than $M_{\rm p}$ then the system has low-$q$.

Because the system masses are drawn from a log-normal 
  distribution then the number
of systems of different masses changes depending on which side of the
peak the systems are found.

For a brown dwarf primary binary to
have a low-$q$ it must come from a lower-mass system than a binary
with high-$q$.  If the primary mass is $M_{\rm p} = 0.06$\msun,
then systems with $q<0.5$ must come from systems of mass 
0.06--0.09\msun, and if $q>0.5$ it must come from systems of
mass 0.09--0.12\msun.  The Chabrier (2003) system IMF has a peak at
about 0.2\msun and so 
there are fewer 0.06--0.09\msun systems than 0.09--0.12\msun
systems and so it would be expected that high-$q$ brown dwarf systems
are more likely than low-$q$.

The reverse argument holds for A-star binaries.  For a 2.5\msun
primary to be in a $q<0.5$ system, the mass of the system must be
2.5--3.75\msun.  Such system masses are more common than the
3.75--5\msun systems a $q>0.5$ binary must form from.

Therefore the {\em mass ratio distributions selected by primary mass are different
to the underlying system mass ratio distribution}.

\bigskip

As can be seen in fig.~\ref{alphazero}, a uniform flat system mass
ratio distribution does a good job of explaining the mass ratio
distributions by primary mass of the stars 
($M_{\rm p} > 0.1$\msun).  M-dwarfs, being near the peak of the 
system mass function,
retain the flat form of the system mass ratio distribution as is
observed (Janson et al. 2012).  G-dwarfs are slightly biased 
towards more unequal-mass
systems (in between the flat and $\alpha=-0.4$ distributions from
Raghavan et al. (2010) and Mentchev \& Hillenbrand (2009)
respectively).  And A-stars have a mass ratio distribution close to
$\alpha = -0.4$ (the dashed-line in fig.~\ref{alphazero}, found by 
Kouwenhoven et al. 2007 and de Rosa et al. 2012).

However, a single flat system mass ratio distribution fails completely
to fit the brown dwarf mass ratio distribution with $\alpha \sim +1.5$
(the dotted line in fig.~\ref{alphazero}, see VLMB archive data).
{\em No
universal system mass ratio distribution can fit both the brown dwarfs
and the stars at the same time.}  

It is worth considering the mass ratio distribution of companions
to massive stars.  In a model with a universal flat mass ratio
distribution B- and O-stars will have a mass ratio distribution
tending to low-mass companions very similar to that of A-stars (for
the same reason).  This
appears at odds with observations of high mass ratios for massive
binaries (e.g. Sana et al. 2012 and references therein).  This could
suggest different binary formation (possibly in a similar way to brown
dwarfs?).  However, observational selection effects mean that
low-mass or distant companions to O-stars are extremely difficult to
detect, making drawing any conclusions from the current data extremely
difficult.

\subsection{The meaning of a flat mass ratio distribution}

It is worth considering what a flat mass ratio distribution actually
means for the distribution of mass in a system.  

If one considers mass
ratios by primary mass, then a flat mass ratio distribution simply
means that the companion is equally likely to have any mass $\leq
M_{\rm p}$.

However, within a system the meaning of a flat mass ratio distribution
is slightly more involved.  The primary must (by definition) have a
mass $M_{\rm p} \geq M_{\rm sys}/2$.  A flat mass ratio distribution
means that the primary is equally likely to have any mass $M_{\rm
  sys}/2 \geq M_{\rm p} > M_{\rm sys}$.  It does {\em not} mean that the
mass is randomly distributed between the two components (and whichever
is the most massive is then the primary).  If the mass is randomly
distributed then low-$q$ systems are more likely (as a 0.9--0.1 split
and a 0.1--0.9 split are equally likely, the distribution is roughly
$f(q) \propto q^{-0.5}$).  We will consider how mass might be
distributed between stars in a system in a later paper.

\section{Conclusions}

We have examined the mass ratio distributions of stars and brown
dwarfs.  We have chosen to examine the mass ratio distributions 
as they are the least likely to have been altered by any dynamical
processing (see Parker \& Goodwin 2012).

We have shown that the different mass ratio distributions of {\em
  stars} when selected by primary mass can all be explained by a
universal, flat, system mass ratio distribution.  However, the
mass ratio distribution of brown dwarfs is very different and cannot
be explained by a universal system mass ratio distribution.

This result would seem to suggest that all binary {\em star} systems (at 
least from 0.2 to 6\msun) select their mass ratio distributions
in the same way -- ie. that the same physical process(es) act to
decide how mass is distributed between the components in a binary.  It
also suggests that this process is very different in low-mass systems
and results in far more equal-mass systems for some reason.  We might
speculate that brown dwarf formation (or at least the formation of
binary brown dwarfs) is fundamentally different to
that of stars in some way (Whitworth et al. 2007; Thies \& Kroupa
2007).

The details of the results will change if a different form 
of the system mass
function is taken.  It is always true that a non-constant system mass
function will change the form of the mass ratio distribution by
primary mass from the underlying distribution and so the conclusions
of this paper are valid unless one believes that system masses are
drawn from a flat distribution.  The underlying mass ratio
distribution is always recovered for system/primary masses near the peak of
the distribution (M-dwarfs for the Chabrier system mass function).  It
is difficult to change the system mass function significantly from the
Chabrier form and recover a standard IMF.

It is important to keep in mind a number of caveats to this work.

Firstly, we have ignored binary fractions and just considered systems
that form a binary system.  This is for the good reason that binary
fractions can change with time and depending on environment, however
to understand star formation we obviously need to understand how and
why some systems fragment into multiple systems whilst others do not,
and how many systems are destroyed by dynamics.

Secondly, we have ignored higher-order multiplicity.  Much recent
evidence suggests that triples and even higher-order multiples are far
more common than once thought (e.g. Tokovinin et al. 2006; 
Eggleton \& Tokovinin 2008; Peter et al. 2012).  Indeed, Peter et al. (2012)
find that half of their A-star multiples are triples (and that is
probably a lower-limit).  How triples and higher-order multiples fit
into the picture of multiple formation is still very unclear.

Thirdly, we have taken the observations at face-value.  Different
samples have different selection effects and biases and it is unclear
how these should be dealt with (see e.g. King et al. 2012a,b).

However, even with these caveats, it is clear that we can have
different interpretations of binary properties depending on if we
group systems by primary mass or system mass.  We argued above that
system mass is probably the more physical mass to take -- even if it
is observationally much more difficult to produce a systematic survey
by system mass (although Gaia should produce such a survey).  And 
we have shown that if we
do sample by system mass, then different masses of (stellar) system 
might well all have the same underlying mass ratio distribution.

%%%%%%%%%%%%%%%%%%%%%%%%%%%%%%%%%%%%%%%%%%%%%%%%%%%%%%%%%%%%%%%%%%%%%%
\section*{Acknowledgements}

Part of this work was done at the International Space Science
Institute (Bern, Switzerland) as part of an International Team.

\label{lastpage}

\end{document}